# RoboRun: A gamification approach to control flow learning for young students with TouchDevelop

Siri Vinay
Department of Computer Science
University College London
siri.vinay.12@ucl.ac.uk

Manoj Vaseekharan
Department of Computer Science
University College London
manoj.vaseekharan.12@ucl.ac.uk

Dean Mohamedally
Department of Computer Science
University College London
d.mohamedally@cs.ucl.ac.uk

## Abstract

This demo paper introduces young students to writing code in a touch enabled interactive maze game. Problem-based learning is given a gamified approach to learning, while simultaneously introducing the TouchDevelop [1] platform to build basic first control flow algorithms and to learn about ordering and loops in conditional statements.

**Categories and Subject Descriptors** K.3.2 [Computer and Information Science Education]: Computer science education

## 1. Introduction

### 1.1 Game idea

RoboRun is an interactive game developed for Windows 8 tablets and touch input devices to teach students basic conditional programming and algorithm sequence ordering skills. RoboRun acts as a basic introduction to coding using a gamification [2] interface to develop key skills and challenge players to come up with better solutions to increasingly more difficult levels. Structuring our application around a game format allows for it to appeal to and engage our target audience of late primary and early secondary school students aged 7 to 14 [3]. Like prior art such as Logo's turtle [4] it uses iconic representation to give immediate visual feedback for debugging with the aim of improving mathematical and algorithmic capability.

### 1.2 RoboRun's Goals

RoboRun aims to inspire children to learn to program through the motivation of gaming by programming a robot for points and challenges. The game lets students experience coding skills needed such as problem solving, basic algorithms and reasoning.

The game enables a student to gain experience in the TouchDevelop language through RoboRun's exercises, which they can then further use to improve their programming ability in the main TouchDevelop environment.

### 1.3 Gameplay

The student controls a robot in a maze using several commands to navigate across the level to a goal point without collisions. There are imperative controls that issue commands such as 'Go Straight','Turn Left', 'Turn Right'. There are also loop statement controls (For loop and While loop) and a decision making statement control (If...else statement). These are directly mapped to TouchDevelop's own 'if statement', 'while loop' and 'for loop' controls.

The player interacts with our game code editor via the use of the touch screen buttons located in the 'controls' section and inputting parameters via dropdown box selections and text inputs. A playback button enables the robot to execute the code in real-time. The student can see whether the robot runs all the way to the goal point, or whether it fails and crashes into an obstacle or wall, with a walkthrough of the code highlighted as it traverses the maze.

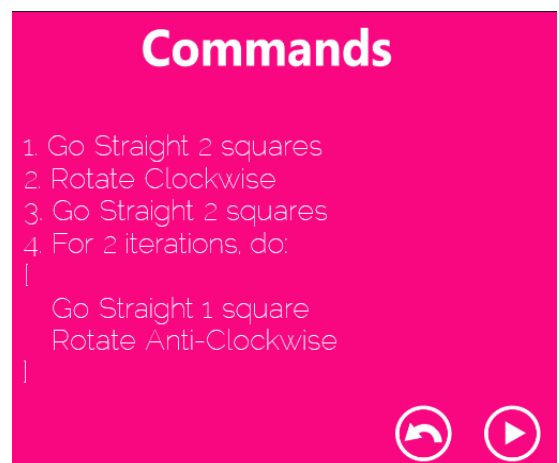

**Figure 1: Structure of pseudo-code syntax in RoboRun**

RoboRun uses full English sentences similar to Scratch [5] to describe the sequence that the robot must follow (figure 1). This allows the solution to be easier to understand by students, as well as forming a link between the Standard English language and a programming language.

The game scores the student player based on various factors such as code length, use of loops/decision statements and time taken to create the solution. Students are discouraged from using solely imperative statements. Displaying a score encourages students to do the level again and obtain a higher score by coming up with more sophisticated solutions.

### 1.4 TouchDevelop Integration with RoboRun

Once a student completes a maze, they are presented with the TouchDevelop generated code enabling them to examine and walkthrough the syntax. Removing the scoring element, this code is then exported to an external TouchDevelop library with the same graphical functionality to be shared on the TouchDevelop cloud service with other students.

Students are able to modify the generated code from the game directly in the TouchDevelop environment. Students can then make use of further programming features from the TouchDevelop platform.

## 2. Design

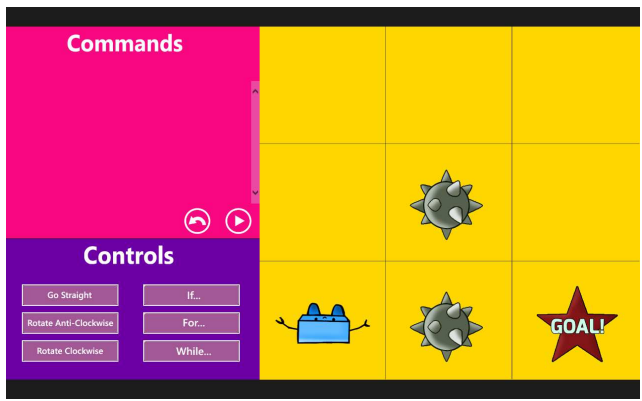

**Figure 2: Gameplay Screen**

The game screen has 3 sections. Figure 2 shows the main section on the right showing the gameplay area with the robot, grid, obstacles and goal point displayed. The top section displays all the commands inputted by the student in a queue. The bottom displays all the controls that the user can use when creating their solution to the level. Controls bring up prompts which the user will have to enter parameters for. For example, using the Go Straight function will ask for the player to enter how many squares they want to move by.

The game was primarily designed for touch enabled/tablet devices. Large square buttons were used for the controls for easy use with arcade like responses. Bright colourful prompts with large circular buttons/input areas, together with vibrant visuals make the game more aesthetically pleasing.

Validation methods prevent the user from entering irrelevant or incorrect data. This indicates to the student that they are perhaps out of bounds and are giving an incorrect command choice.

## 3. Upcoming Trial Work

The game will be trialled in classroom settings with students of a range of age groups similar to previous e-learning approaches [6]. User centred evaluations will report time to solve algorithms of different approaches, also considering the students background experience to programming and how they collaborate with other students to solve the mazes. A comparison of approaches will be suggested for improving control flow teaching with modern devices.

Several levels and game tasks are currently being implemented including a "Build your own level" feature to design and test a student's ability to create longer algorithms for solving mazes. A planned follow up development will incorporate additional gameplay mechanics for the robot such as wait-for states for obstacles, swapping data to teleport, and multidimensional arrays of positions in a 3D view of the gameplay.